\documentclass[12pt]{article}
\usepackage{epsfig}

\newcommand{\mysection}{\setcounter{equation}{0}\section}

\def\beq{\begin{equation}}
\def\eeq{\end{equation}}
\def\beqa{\begin{eqnarray}}
\def\eeqa{\end{eqnarray}}
 
\newlength{\dinwidth} \newlength{\dinmargin}
\begin{document}

\begin{center}
{\Large \bf The top quark rapidity distribution and forward-backward asymmetry}
\end{center}
\vspace{2mm}
\begin{center}
{\large Nikolaos Kidonakis}\\
\vspace{2mm}
{\it Kennesaw State University,  Physics \#1202,\\
1000 Chastain Rd., Kennesaw, GA 30144-5591, USA}
\end{center}
 
\begin{abstract}
I present results for the top quark rapidity distribution at LHC and Tevatron 
energies, including higher-order corrections from threshold resummation.
The next-to-next-to-leading-order (NNLO) soft-gluon corrections at 
next-to-next-to-leading-logarithm (NNLL) level are added to the NLO result.
Theoretical predictions are shown for the rapidity distribution, 
including the scale dependence of the distributions. 
The forward-backward asymmetry at the Tevatron is also calculated.
\end{abstract}
 
\mysection{Introduction}

The study of the top quark has a central role in current collider physics 
research programs. 
The experimental measurements of the $t{\bar t}$ quark cross section at 
the Tevatron \cite{CDFcs,D0cs} and  the LHC \cite{ATLAS,CMS}, and of the 
top quark transverse momentum distribution 
at the Tevatron \cite{CDFpT,D0pT}
are currently in good agreement with theoretical predictions 
\cite{NKttbar,NKDIS2011}. 
The rapidity distribution and the forward-backward asymmetry 
(or charge asymmetry) have also been measured at the Tevatron 
\cite{CDFasym,D0asym}. 
The forward-backward asymmetry has been found to be surprisingly large.
This apparent discrepancy with theory, as well as the fact that experimental 
errors 
continue to get smaller, make precise theoretical calculations in the Standard 
Model necessary, in order to be able to clearly identify any effects of 
new physics.

Next-to-leading order (NLO) calculations of the QCD corrections to 
$t {\bar t}$ production have been available for over 
two decades \cite{NLO1,NLO2}  
but the associated uncertainty is much bigger than current experimental 
errors for the total cross section. 
The inclusion of higher-order soft-gluon corrections enhances the cross 
section and transverse momentum distribution and significantly 
reduces the theoretical error \cite{NKttbar}. 

Recent theoretical predictions use approximate next-to-next-to-leading order 
(NNLO) calculations based on next-to-next-to-leading-logarithm (NNLL) 
resummation of soft-gluon corrections for the differential cross section 
\cite{NKttbar}. 
To achieve NNLL accuracy in the 
resummation we have calculated the soft anomalous dimensions at two loops 
\cite{NKttbar,NK2l,NKsingt}.
These soft-gluon corrections dominate the cross section for $t{\bar t}$ 
production and at first-order they provide an excellent 
approximation to the exact NLO corrections at both Tevatron and LHC 
energies \cite{NKttbar,NKDIS2011}. 

We begin with the double differential cross section, 
$d^2\sigma/(dp_T^2 \, dY)$, where $p_T$ is the 
transverse momentum of the top quark, and $Y$ is the rapidity of the top 
quark in the hadronic center-of-mass frame.   
We use our resummation formalism to calculate soft-gluon contributions 
for this differential cross section
(see \cite{NKttbar} and references therein). The total cross section was 
calculated in \cite{NKttbar} by integrating over both $p_T$ and rapidity.
In \cite{NKttbar} the $p_T$ distribution, $d\sigma/dp_T$ was also calculated 
by integrating the double differential cross section over rapidity.

In this paper we calculate the rapidity distribution, $d\sigma/dY$, 
by integrating 
the double differential cross section over the transverse momentum
\beq
\frac{d\sigma}{dY}= \int_0^{p_{T^+}^2} \, 
\frac{d^2\sigma}{dp_T^2 \, dY} \, dp_T^2 
\label{hadropt}
\eeq
where the upper limit of integration is 
$p_{T^+}^2=S/(4 \cosh^2 Y)-m^2$, with $m$ the top quark mass and 
$S$ the squared hadronic center-of-mass energy.
The total cross section can also be calculated by integrating 
Eq. (\ref{hadropt}) over $Y$ with integration 
limits $\pm (1/2) \ln[(1+\beta)/(1-\beta)]$ 
where  $\beta=\sqrt{1-4m^2/S}$, which serves as a further check on the 
calculation.

In the next Section, we calculate the rapidity distribution at the LHC 
at 7 and 14 TeV energy while in Section 3 we do the calculation for 
Tevatron energy. In Section 4 we discuss the top quark forward-backward 
asymmetry at the Tevatron. We conclude in Section 5. 

\mysection{Top quark rapidity distribution at the LHC}

We begin with a study of the top quark rapidity distribution 
at the LHC. We show results for the current LHC energy of 7 TeV 
and the future (design) energy of 14 TeV.
We present NLO and approximate NNLO calculations for the rapidity distribution. 
The NNLO approximate rapidity distribution is computed by adding the NNLO soft-gluon 
corrections, derived from NNLL resummation, to the exact NLO result.
In our calculations we use the MSTW2008 NNLO parton distribution functions 
\cite{MSTW2008}.

\begin{figure}
\begin{center}
\includegraphics[width=11cm]{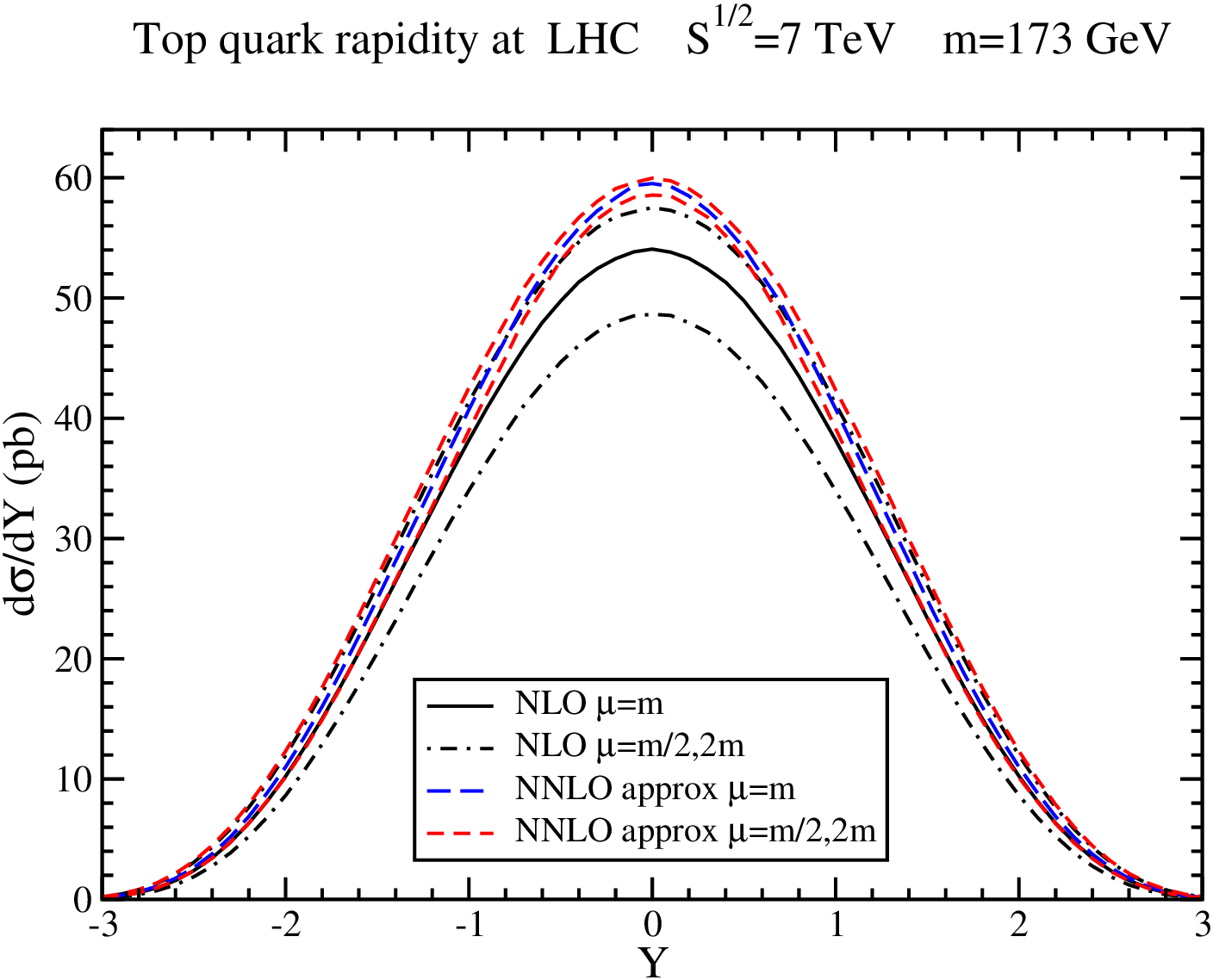}
\caption{The top quark rapidity distribution  
at the LHC with $\sqrt{S}=7$ TeV, $m=173$ GeV, 
and $\mu=m/2$, $m$, $2m$.}
\label{y7lhclin}
\end{center}
\end{figure}

\begin{figure}
\begin{center}
\includegraphics[width=11cm]{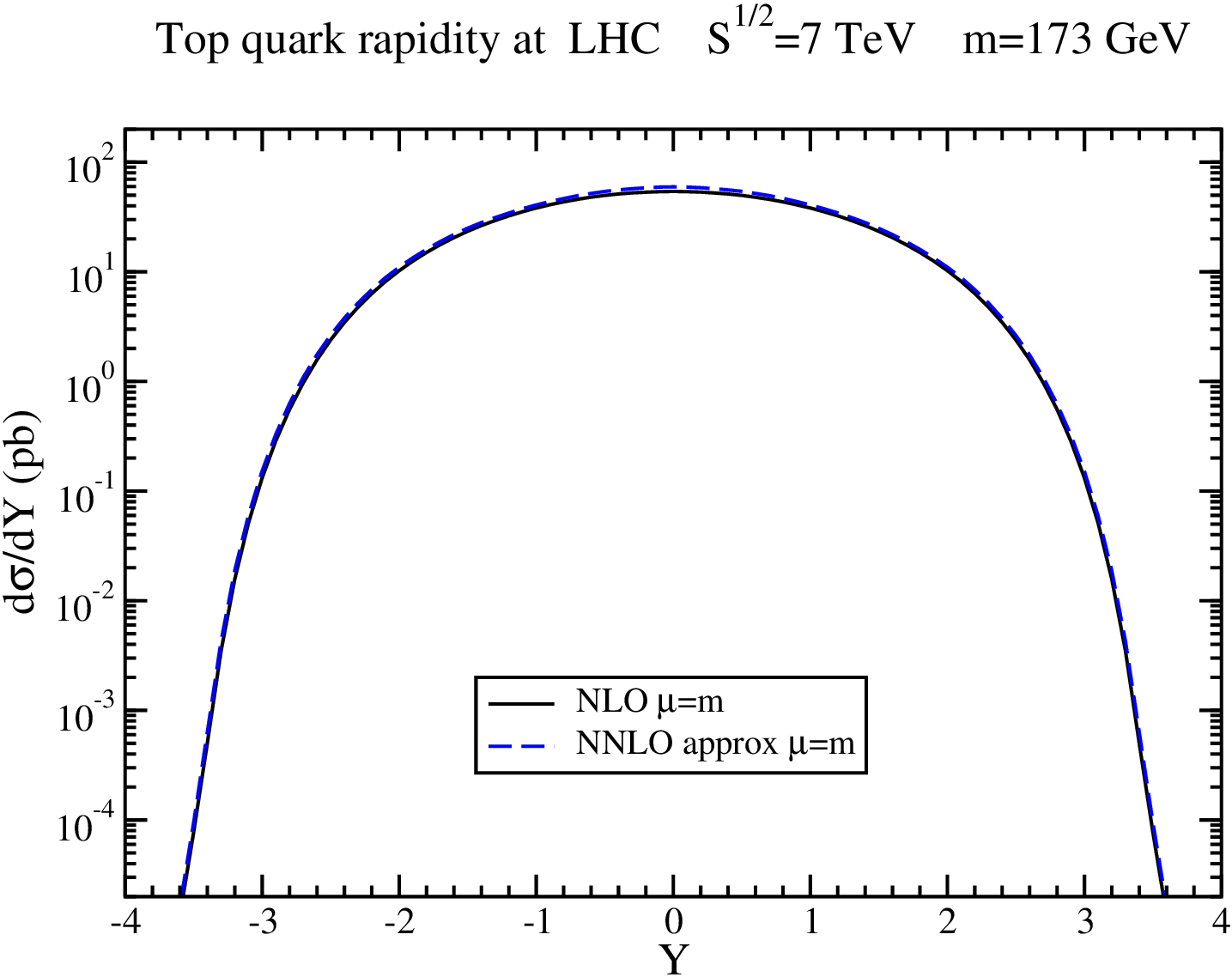}
\caption{The top quark rapidity distribution  
at the LHC with $\sqrt{S}=7$ TeV and $\mu=m=173$ GeV in a logarithmic plot.}
\label{y7lhclog}
\end{center}
\end{figure}

The top quark rapidity distribution  
at the LHC at 7 TeV energy is plotted 
in Figs. \ref{y7lhclin} and \ref{y7lhclog}. 
We use $m=173$ GeV, currently the best 
value for the top quark mass \cite{TevEWG}.
We denote by $\mu$ the common factorization and renormalization scale.  
Fig. \ref{y7lhclin} shows NLO and approximate NNLO results for 
the differential distribution $d\sigma/dY$ 
for three different scale choices, 
$\mu=m/2$, $m$, and $2m$. 
The scale variation of the $Y$ distribution at NNLO is much  
smaller than that at NLO, consistent with the results in Ref. \cite{NKttbar} 
for the total cross section and $p_T$ distribution. 

Figure \ref{y7lhclog} presents the results at 7 TeV
for  $d\sigma/dY$ in a logarithmic plot for a wider range of rapidity values.
It is clear that $d\sigma/dY$ falls off very quickly for larger rapidities.  
From both Figs. \ref{y7lhclin} and \ref{y7lhclog} we see that the 
NNLO soft-gluon corrections contribute an 
enhancement to  the NLO rapidity distribution, but the shapes of the 
NLO and approximate NNLO distributions are similar.

\begin{figure}
\begin{center}
\includegraphics[width=11cm]{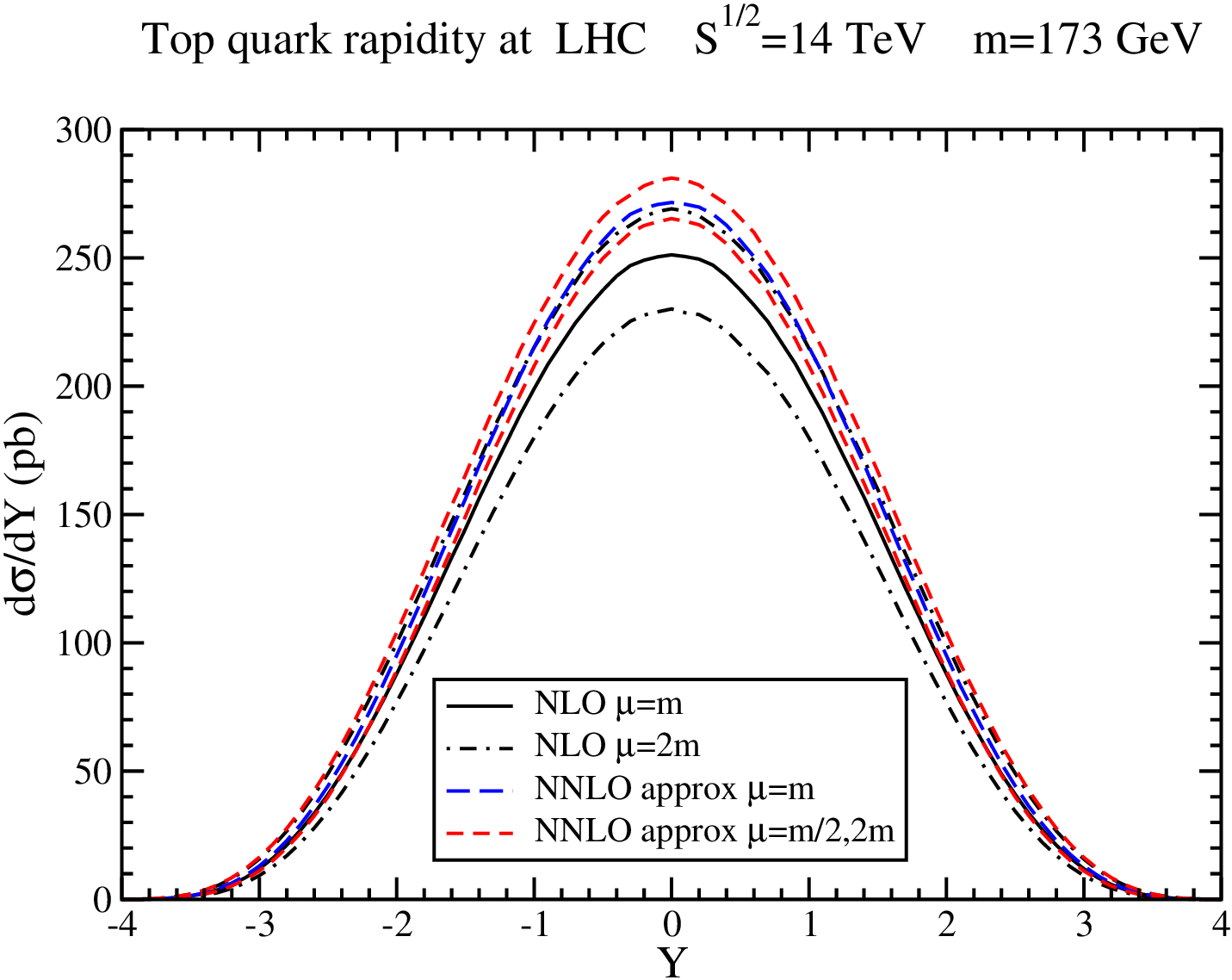}
\caption{The top quark rapidity distribution  
at the LHC with $\sqrt{S}=14$ TeV, $m=173$ GeV, and $\mu=m/2$, $m$, $2m$.}
\label{y14lhclin}
\end{center}
\end{figure}

\begin{figure}
\begin{center}
\includegraphics[width=11cm]{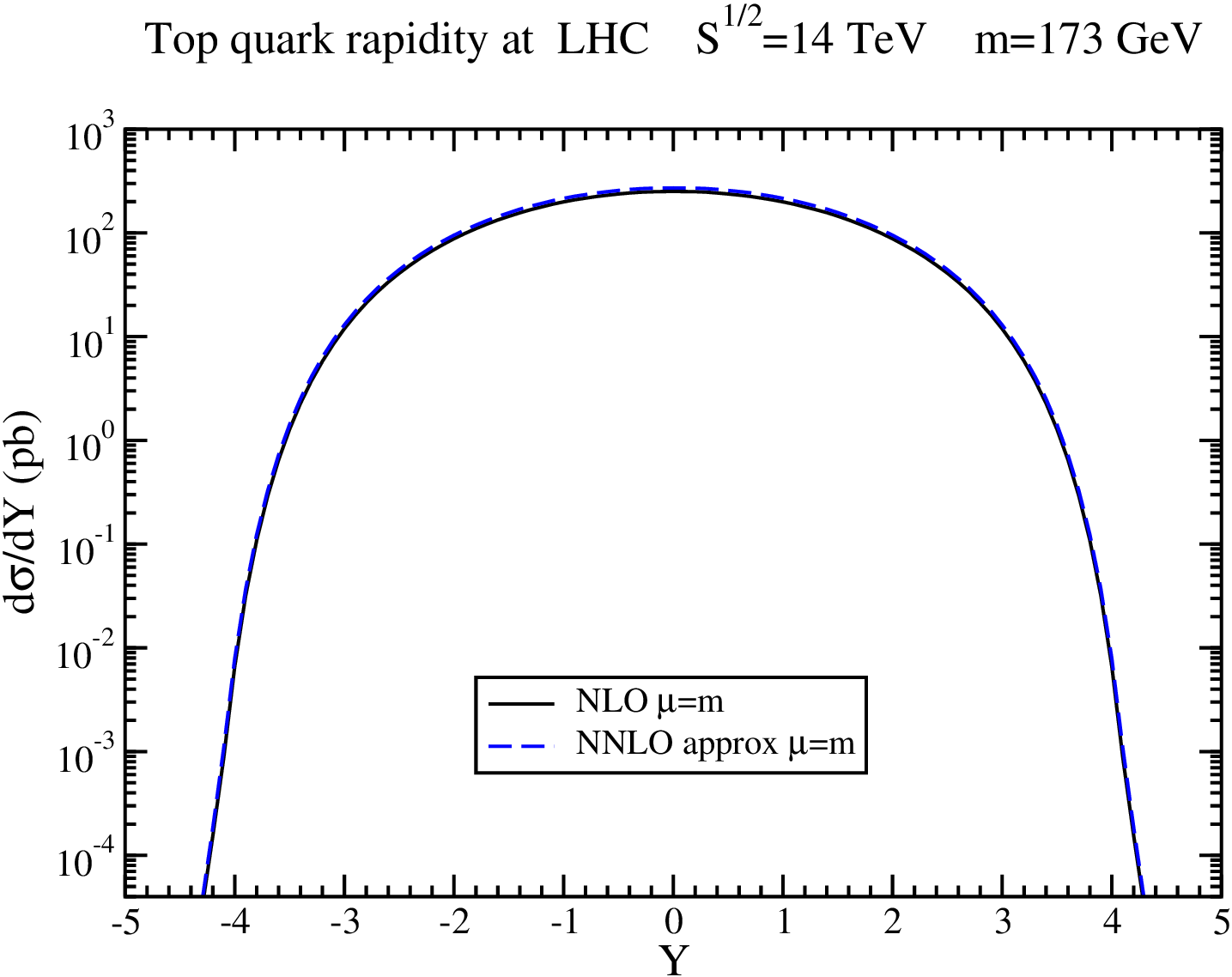}
\caption{The top quark rapidity distribution  
at the LHC with $\sqrt{S}=14$ TeV and $\mu=m=173$ GeV in a logarithmic plot.}
\label{y14lhclog}
\end{center}
\end{figure}

The rapidity distribution of the top quark with $m=173$ GeV  
at the LHC at 14 TeV energy is plotted 
in Figs. \ref{y14lhclin} and \ref{y14lhclog}. 
Fig. \ref{y14lhclin} shows NLO and approximate NNLO results 
for three different scale choices, 
$\mu=m/2$, $m$, and $2m$. 
Again, the scale variation of the $Y$ distribution at NNLO is much  
smaller than that at NLO.  

Figure \ref{y14lhclog} presents the results for  $d\sigma/dY$ at 14 TeV 
in a logarithmic plot for a wider range of $Y$ values. The rapidity ranges 
shown in Figs. \ref{y14lhclin} and \ref{y14lhclog} are of course wider 
than the corresponding ones in Figs. \ref{y7lhclin} and \ref{y7lhclog}, 
since the rate increases significantly at the higher energy.
At 14 TeV the NNLO soft-gluon corrections provide a 
significant contribution, but the shapes of the NLO and the approximate NNLO 
distributions are similar.

In all four figures we see that the rapidity distributions at the LHC are 
fairly symmetric.
This is due to the fact that the $gg \rightarrow t{\bar t}$ channel is dominant
at the LHC, and this channel is completely symmetric at all orders. 
We discuss this in more detail in Section 4. 

Finally, we note that the integrated rapidity distribution gives the same 
result for the total cross section as found in \cite{NKttbar} at both LHC 
energies, which provides a good consistency check of the calculation. 

\mysection{Top quark rapidity distribution at the Tevatron}

We continue with the top quark rapidity distribution at the Tevatron collider.
Again, we present exact NLO and approximate NNLO (from NNLL resummation)  
results. 

\begin{figure}
\begin{center}
\includegraphics[width=11cm]{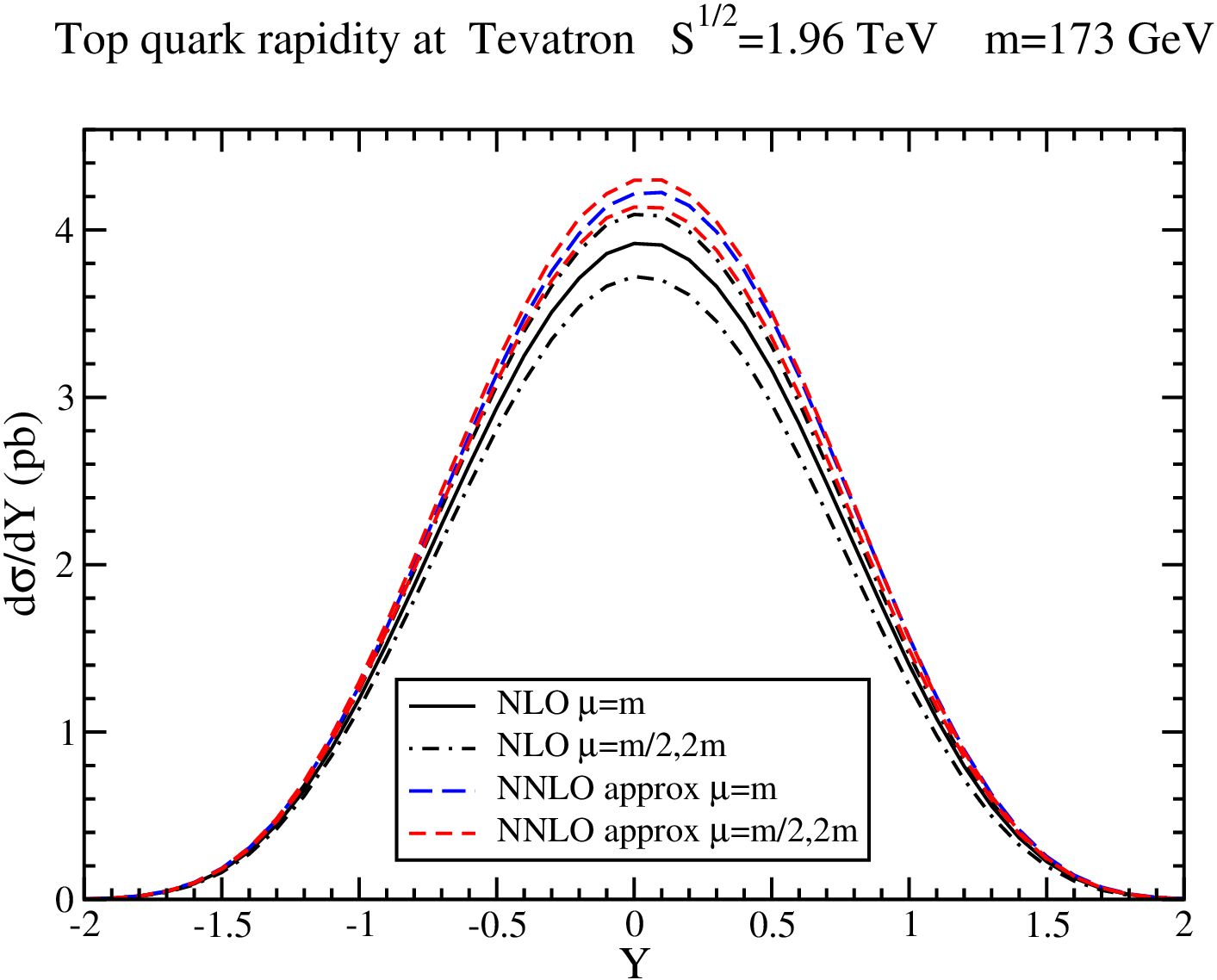}
\caption{The top quark rapidity distribution  
at the Tevatron with $\sqrt{S}=1.96$ TeV, $m=173$ GeV, 
and $\mu=m/2$, $m$, $2m$.}
\label{ytevlin}
\end{center}
\end{figure}

\begin{figure}
\begin{center}
\includegraphics[width=11cm]{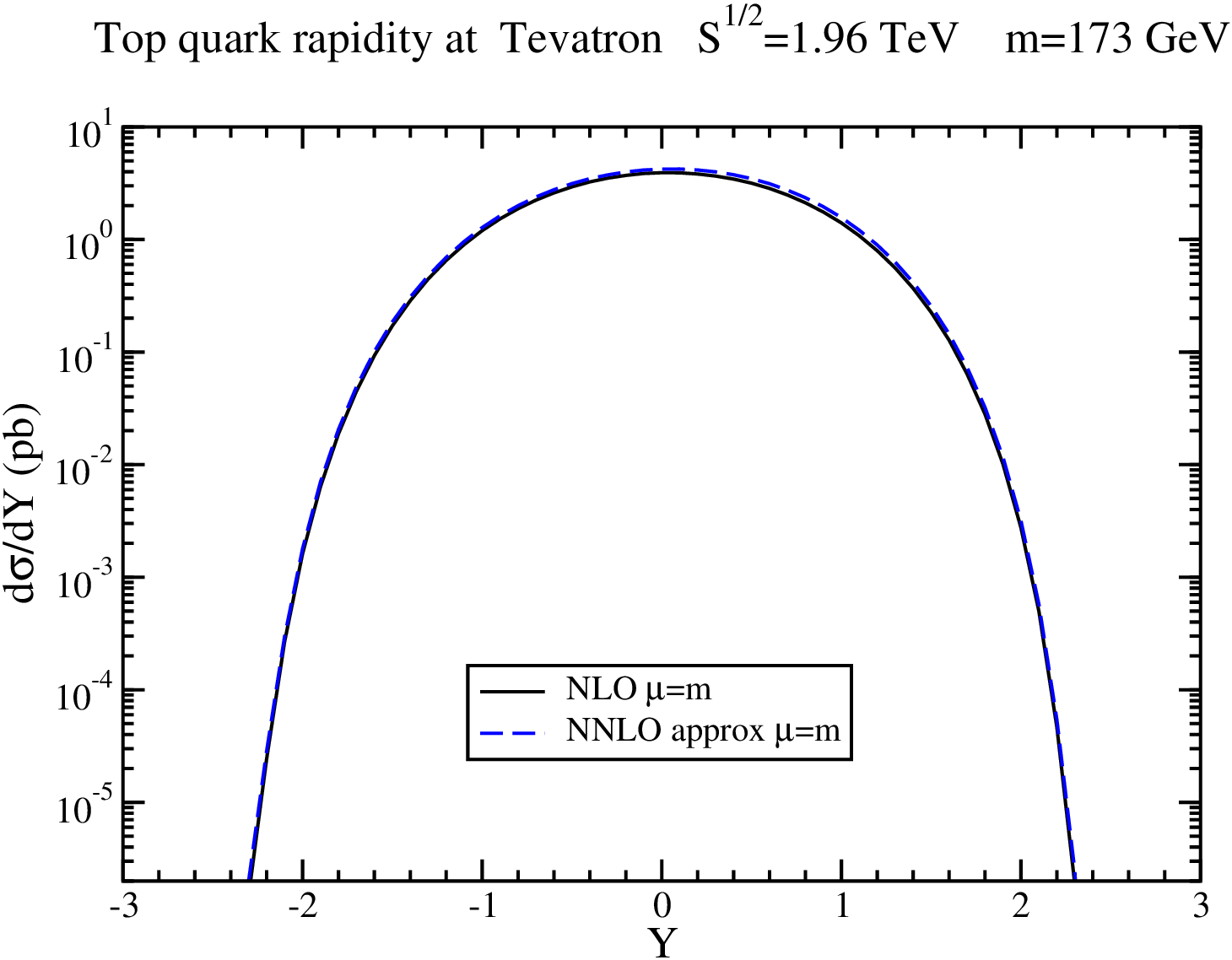}
\caption{The top quark rapidity distribution  
at the Tevatron with $\sqrt{S}=1.96$ TeV and $\mu=m=173$ GeV 
in a logarithmic plot.}
\label{ytevlog}
\end{center}
\end{figure}

The top quark rapidity distribution at the Tevatron 
with $m=173$ GeV is plotted in Figs. \ref{ytevlin} and \ref{ytevlog}. 
Fig. \ref{ytevlin} shows the differential distribution $d\sigma/dY$ at 
both NLO and approximate NNLO for three different scale choices, 
$\mu=m/2$, $m$, and $2m$. 
The integrated rapidity distribution gives the total cross 
section found in \cite{NKttbar}, which is a good consistency 
check of the calculation. The scale variation of the 
$Y$ distribution at NNLO is significantly smaller than at NLO, again 
as also found for the total cross section and the $p_T$ distribution 
in \cite{NKttbar}.  
The NNLO soft-gluon corrections enhance the NLO result but the shape is 
not significantly affected.

Figure \ref{ytevlog} presents the top quark rapidity distribution at the 
Tevatron in a logarithmic plot that makes it easier to see $d\sigma/dY$ 
for larger $Y$ values.
Again the fall of the distribution at larger $Y$ is very steep.

We note that, unlike the LHC results, at the Tevatron the rapidity 
distribution 
of the top quark is clearly not symmetric. The maximum of the distribution 
is not at $Y=0$ but at positive $Y$. We discuss this forward-backward 
asymmetry in the next Section. 

\mysection{Top quark forward-backward asymmetry at the Tevatron}

We define the top quark forward-backward asymmetry as 
\beq
A_{\rm FB}=\frac{\sigma(Y>0)-\sigma(Y<0)}{\sigma(Y>0)+\sigma(Y<0)} \, .
\eeq

The asymmetry has been calculated in \cite{HHK,KR,BER} in NLO QCD, 
and more recently 
using threshold resummation at NLL in \cite{ASV} and using SCET at 
NNLL in \cite{AFNPY}.

The leading-order (LO) production channels, $q{\bar q} \rightarrow t{\bar t}$ 
and $gg \rightarrow t{\bar t}$, are symmetric in rapidity, thus 
$A_{FB}$ vanishes at LO. The $gg$ channel remains symmetric at all orders. 
However an asymmetry 
arises in the $q{\bar q}$ channel starting at NLO. Furthermore asymmetry 
arises from flavor excitation, $q g \rightarrow q t {\bar t}$ 
\cite{NLO1,NLO2,HHK,KR,BER}.

Therefore by applying resummation we expect the $gg$ channel to remain 
symmetric, but we will have contributions to the asymmetry from higher orders 
in the $q{\bar q}$ channel. Since the $gg$ channel is dominant at the LHC, 
 the asymmetry there is very small, as can also be seen from the 
fairly symmetric rapidity distributions presented in Section 2.
At the Tevatron, on the other hand, the $q{\bar q}$ channel is dominant and 
the asymmetry is larger and evident from the rapidity distributions presented 
in Section 3.

The measurements of $A_{\rm FB}$ at CDF \cite{CDFasym} and D0 \cite{D0asym}
have returned values substantially larger than the Standard Model prediction,
so it is important to perform the most accurate calculation to have 
confidence in the theoretical prediction while seeking hints of new physics.

Using the NNLO approximate rapidity distributions in the previous section, 
we find a top quark forward-backward asymmetry at the Tevatron of 0.052, 
or 5.2\%, in the $p{\bar p}$ center-of-mass frame. 
This is a modest increase on the 4\% NLO asymmetry. 
Uncertainties on this number can be estimated by varying the 
scale $\mu$ between $m/2$ and $2m$ as shown for the rapidity distributions.
The value of 0.052 found for $\mu=m$ is a maximum, but the number can vary 
down to 0.046 for $\mu=2m$, so we write 
$A_{\rm FB}=0.052^{+0.000}_{-0.006}$.
Current measurements at the Tevatron indicate asymmetries of 15\% or more, 
with around two standard deviations excess over the theoretical prediction. 

\mysection{Conclusions}

We have shown in this paper that the top quark rapidity distributions 
at the LHC and the Tevatron receive  
significant enhancements from soft-gluon corrections. These corrections have 
been resummed at NNLL accuracy by using the two-loop soft anomalous dimension 
matrices for the partonic processes.
Approximate NNLO rapidity distributions have been derived from 
the NNLL resummed result. The NNLO soft-gluon corrections enhance the  
top quark rapidity distribution and greatly reduce the theoretical 
uncertainty from scale variation.
The top quark forward-backward asymmetry at the Tevatron has been calculated. 
The theoretical prediction of 5.2\% is significantly smaller than current 
experimental values. 

\mysection*{Acknowledgements}
This work was supported by the National Science Foundation under 
Grant No. PHY 0855421.


\begin{thebibliography}{99}

\bibitem{CDFcs}
CDF Collaboration, T. Aaltonen {\it et al.}, 
Phys. Rev. D {\bf 79}, 052007 (2009) [arXiv:0901.4142 [hep-ex]]; 
Phys. Rev. D {\bf 79}, 112007 (2009) [arXiv:0903.5263 [hep-ex]];  
Phys. Rev. D {\bf 81}, 052011 (2010) [arXiv:1002.0365 [hep-ex]];
Phys. Rev. D {\bf 82}, 052002 (2010) [arXiv:1002.2919 [hep-ex]];  
Phys. Rev. D {\bf 81}, 092002 (2010) [arXiv:1002.3783 [hep-ex]];
Phys. Rev. D {\bf 83}, 071102 (2011) [arXiv:1007.4423 [hep-ex]]; 
arXiv:1103.4821 [hep-ex]; arXiv:1105.1806 [hep-ex];
Conf. Note 9913.  

\bibitem{D0cs}
D0 Collaboration, V.M. Abazov {\it et al.}, 
Phys. Rev. Lett. {\bf 100}, 192003 (2008) [arXiv:0801.1326 [hep-ex]]; 
Phys. Rev. Lett. {\bf 100}, 192004 (2008) [arXiv:0803.2779 [hep-ex]]; 
Phys. Lett. B {\bf 679}, 177 (2009) [arXiv:0901.2137 [hep-ex]]; 
Phys. Rev. D {\bf 80}, 071102 (2009) [arXiv:0903.5525 [hep-ex]]; 
Phys. Rev. D {\bf 82}, 032002 (2010) [arXiv:0911.4286 [hep-ex]]; 
Phys. Rev. D {\bf 82}, 071102 (2010) [arXiv:1008.4284 [hep-ex]];
arXiv:1101.0124 [hep-ex]; arXiv:1104.2887 [hep-ex].

\bibitem{ATLAS}
ATLAS Collaboration, G. Aad {\it et al.}, Eur. Phys. J. C {\bf 71}, 1577 (2011)
[arXiv:1012.1792 [hep-ex]]; 
ATLAS-CONF-2011-040; ATLAS-CONF-2011-054.

\bibitem{CMS}
CMS Collaboration, V. Khachatryan {\it et al.}, Phys. Lett. B {\bf 695}, 424 (2011) [arXiv:1010.5994 [hep-ex]];
CMS-PAS-TOP-11-001.

\bibitem{CDFpT}
CDF Collaboration, A.A. Affolder et al., Phys. Rev. Lett. {\bf 87}, 102001 (2001); CDF II Collaboration, CDF Note 10234. 

\bibitem{D0pT}
D0 Collaboration, V.M. Abazov {\it et al.}, Phys. Lett. B {\bf 693}, 515 (2010)
[arXiv:1001.1900 [hep-ex]]. 

\bibitem{NKttbar}
N. Kidonakis, Phys. Rev. D {\bf 82}, 114030 (2010) [arXiv:1009.4935 [hep-ph]].

\bibitem{NKDIS2011}
N. Kidonakis, in {\sl DIS 2011}, arXiv:1105.3481 [hep-ph].

\bibitem{CDFasym}
CDF Collaboration, T. Aaltonen {\it et al.}, Phys. Rev. Lett. {\bf 101}, 
202001 (2008) [arXiv:0806.2472 [hep-ex]]; CDF Note 9724; 9853; 10224; 10436;
arXiv:1101.0034 [hep-ex].

\bibitem{D0asym}
D0 Collaboration, V.M. Abazov {\it et al.}, Phys. Rev. Lett. {\bf 100}, 
142002 (2008) [arXiv:0712.0851 [hep-ex]]; D0 Note 6062-CONF. 

\bibitem{NLO1}
P. Nason, S. Dawson, and R.K. Ellis,
Nucl. Phys. B {\bf 303}, 607 (1988); Nucl. Phys. {\bf 327}, 49 (1989); (E) B {\bf 335}, 260 (1990). 

\bibitem{NLO2}
W. Beenakker, H. Kuijf, W.L. van Neerven, and J. Smith,
Phys. Rev. D {\bf 40}, 54 (1989); \\
W. Beenakker, W.L. van Neerven, R. Meng, G.A. Schuler, and J. Smith,
Nucl. Phys. B {\bf 351}, 507 (1991).

\bibitem{NK2l}
N. Kidonakis, Phys. Rev. Lett. {\bf 102}, 232003 (2009) 
[arXiv:0903.2561 [hep-ph]].

\bibitem{NKsingt}
N. Kidonakis, Phys. Rev. D {\bf 81}, 054028 (2010) [arXiv:1001.5034 [hep-ph]]; 
Phys. Rev. D {\bf 82}, 054018 (2010) [arXiv:1005.4451 [hep-ph]]; 
Phys. Rev. D {\bf 83}, 091503(R) (2011) [arXiv:1103.2792 [hep-ph]].

\bibitem{MSTW2008}
A.D. Martin, W.J. Stirling, R.S. Thorne, and G. Watt, 
Eur. Phys. J. C {\bf 63}, 189 (2009) [arXiv:0901.0002 [hep-ph]].

\bibitem{TevEWG}
Tevatron Electroweak Working Group, arXiv:1007.3178.

\bibitem{HHK}
F. Halzen, P. Hoyer, and C.S. Kim, Phys. Lett. B {\bf 195}, 74 (1987).

\bibitem{KR}
J.H. Kuhn and G. Rodrigo, Phys. Rev. Lett. {\bf 81}, 49 (1998) 
[hep-ph/9802268]; Phys. Rev. D {\bf 59}, 054017 (1999) [hep-ph/9807420].

\bibitem{BER}
M.T. Bowen, S.D. Ellis, and D. Rainwater, Phys. Rev. D {\bf 73}, 014008 (2006) 
[hep-ph/0509267].

\bibitem{ASV}
L.G. Almeida, G. Sterman, and W. Vogelsang, Phys. Rev. D {\bf 78},  
014008 (2008) [arXiv:0805.1885 [hep-ph]].

\bibitem{AFNPY}
V. Ahrens, A. Ferroglia, M. Neubert, B.D. Pecjak, and L.L. Yang, 
arXiv:1103.0550 [hep-ph].

\end{thebibliography}
\end{document}